# Molecular structure refinement by direct fitting of atomic coordinates to experimental ESR spectra

G.T.P. Charnock, M. Krzystyniak, Ilya Kuprov[*]

*Oxford e-Research Centre, University of Oxford,*

*7 Keble Road, Oxford, OX1 3QG, UK.*

Fax: +44 1865 610 612

Email: ilya.kuprov@oerc.ox.ac.uk

**Abstract**

An attempt is made to bypass spectral analysis and fit internal coordinates of radicals directly to experimental liquid- and solid-state electron spin resonance (ESR) spectra. We take advantage of the recently introduced large-scale spin dynamics simulation algorithms and of the fact that the accuracy of quantum mechanical calculations of ESR parameters has improved to the point of quantitative correctness. Partial solutions are offered to the local minimum problem in spectral fitting and to the problem of spin interaction parameters (hyperfine couplings, chemical shifts, *etc.*) being very sensitive to distortions in molecular geometry.

# 1. Introduction

Molecular structure determination using magnetic resonance spectroscopy has reached remarkable levels of accuracy and sophistication – particularly in NMR, it is currently possible to determine atomic coordinates of small proteins in several days or even hours[1-5], courtesy of the dipole-dipole couplings that are directly related to inter-atomic distances[6-9], and *J*-couplings that propagate through the bonding network and depend on dihedral angles[10]. In crystalline materials, the structures can be refined further by fitting atomic coordinates, using electronic structure theory calculations[11-14], to the experimentally determined quadrupolar and chemical shielding tensors[15,16]: the coordinates are varied until the computed magnetic parameters match the experimental ones.

Electron spin resonance structure determination is more complicated due to the fact that ESR parameters (hyperfine couplings and *g*-tensors) depend very indirectly and often quite strongly on the molecular geometry[17,18], a good example being tyrosyl radical, where the hyperfine coupling to the $CH_2$ group protons can be anywhere between zero and 20 Gauss, depending on the dihedral angle with the aromatic ring[19,20]. The tyrosyl radical structure problem has recently been solved, heroically, by scanning every significant degree of freedom[19], but it is clear that a less labour-intensive method would be appreciated.

In this paper we describe and evaluate a procedure wherein the atomic coordinates are iterated directly against experimental ESR data. The algorithm relies on two recent developments in chemical physics: firstly, ESR parameters can now be computed reliably and efficiently from basic principles for medium-sized molecules[11-14,17,18,21], and secondly, a spin dynamics simulation package capable of treating large spin systems in reasonable time has recently emerged[22-24]. It is now possible, by combining the two, to perform direct fitting of atomic coordinates to experimental ESR spectra by iteratively minimizing the deviation of the theoretical spectrum from the experimental one.

The idea of fitting atomic coordinates directly to experimental magnetic resonance data has been around in various forms for several years[15,25-29]: structures were fitted to numerical values of magnetic resonance parameters[15,16,26,29] and to magnetic resonance spectra with some approximations (accurate spectral simulations being prohibitively expensive)[25]. The approach described in this paper is more general and supports the method proposed by Polimeno and co-authors[28] – we use accurate quantum chemical calculations as well as accurate spin dynamics simulations (a schematic is given in Figure 1) with the recently developed large-scale NMR/ESR simulation tools[22-24]. We also employ the L-BFGS quasi-Newton optimization method with line search[30] that significantly reduces the number of geometry iterations at the fitting stage compared to earlier algorithms.

## 2. General formalism

The basic principle of magnetic resonance structure determination can be summarized as follows: *a chemically reasonable structure that exhibits a set of magnetic resonance spectra similar to those experimentally observed is likely to be a good representation of the true structure.* This statement can be formalized as a minimization condition for the following error functional:

$$\Omega(\vec{r}) = \left\| S_{\text{exp}} - S(\mathbf{m}(\vec{r})) \right\|^2 + \lambda E(\vec{r}), \tag{1}$$

where $S_{\text{exp}}$ is the experimental magnetic resonance spectrum and $S(\mathbf{m}(\vec{r}))$ is the theoretical spectrum that depends on a set of magnetic interaction parameters $\mathbf{m} = \{m_1, ..., m_k\}$, which in turn depend on the vector of atomic coordinates $\vec{r}$. $E(\vec{r})$ is the total molecular energy and $\lambda$ is a weighting factor regulating the relative importance of molecular energy *versus* spectral properties. The energy term is responsible for keeping the structure "chemically reasonable" and is also required to constrain the parts of the molecule that do not influence the spectrum.

Using the Hilbert space $L^2$-norm, we get, for an $N$-dimensional magnetic resonance spectrum, the following error functional:

$$\Omega(\vec{r}) = \int \left| S_{\text{exp}} - S(\mathbf{m}(\vec{r})) \right|^2 d\omega_1 ... d\omega_N + \lambda E(\vec{r}). \tag{2}$$

where $\{\omega_1, ..., \omega_N\}$ are the frequency variables. Within the Born-Oppenheimer approximation $\Omega(\vec{r})$ only depends on the nuclear coordinates. Its gradient with respect to $\vec{r}$

$$\nabla \Omega(\vec{r}) = -\int \left[ \left( \sum_k \frac{\partial S}{\partial m_k} \nabla m_k(\vec{r}) \right) \left[ S_{\text{exp}} - S(\mathbf{m}(\vec{r})) \right]^* + \text{c.c.} \right] d\omega_1 ... d\omega_N + \lambda \nabla E(\vec{r}) \tag{3}$$

contains three types of derivatives: energy with respect to coordinates, simulation with respect to magnetic parameters and magnetic parameters with respect to coordinates. The gradient of energy[31] and magnetic resonance simulation[32] are well researched and present little difficulty, but the coordinate derivatives of magnetic interaction parameters require closer attention.

## 3. Derivatives of the error functional

The computational complexity of spin interaction derivatives can be very considerable: the geometric gradient of a *g*-tensor is a third order response property. Their evaluation is further complicated by the fact that quantum chemistry software implementing these derivatives analytically does not at present exist – all calculations presented in this work therefore used the four-point central finite difference approximation:

$$f'(x) = \frac{f(x-2h) - 8f(x-h) + 8f(x+h) - f(x+2h)}{12h} + O(h^4) \tag{4}$$

for gradient evaluation. This is sufficient for the purposes of the present paper, but analytical derivatives of magnetic parameters are very desirable and would accelerate the calculations as well as improve convergence of the error functional minimization procedure. One practical advantage of Equation (4), however, is easy parallelization: for a typical run with 50-100 internal coordinates, it generates several hundred independent DFT jobs that may be evaluated in parallel.

Because the geometry must represent a chemically sensible structure, the energy cost of every step in the atomic coordinate space must be considered explicitly. For a given molecular geometry, an infinitesimal coordinate displacement $d\vec{r}$ would produce a perturbation $dm_k$ of a magnetic parameter $m_k$:

$$dm_k = \nabla m_k \cdot d\vec{r} \tag{5}$$

and produce a change $dE$ in the total energy:

$$dE = \nabla E \cdot d\vec{r}. \tag{6}$$

Therefore, for a given direction vector $\vec{n}$ in the atomic coordinate space and a given infinitesimal displacement $dx$ along that direction, the following quantity

$$C_k(\vec{n}) = \frac{dm_k}{dE} = \frac{\nabla m_k \cdot d\vec{r}}{\nabla E \cdot d\vec{r}} = \frac{\nabla m_k \cdot \vec{n} dx}{\nabla E \cdot \vec{n} dx} = \frac{\nabla m_k \cdot \vec{n}}{\nabla E \cdot \vec{n}} \tag{7}$$

gives the energy cost of perturbation in the magnetic parameter $m_k$ when the geometry is distorted in the atomic coordinate direction given by $\vec{n}$. The coordinate derivatives of hyperfine couplings for a simple example of a gas-phase allyl radical are given in Table 1. It is clear that even minor perturbations of molecular geometry can, for some coordinates, lead to significant perturbations of hyperfine couplings: a 0.2 Gauss change in isotropic HFC for a 0.01 Å change in geometry is typical. The associated energy costs (listed in Table 2) are quite modest – a room temperature energy budget of $RT = 2.48$ kJ/mol is sufficient to perturb some proton hyperfine couplings by as much as 3 Gauss.

This steep geometry dependence of hyperfine couplings (as well as other magnetic couplings) means in practice that, unless full vibrational and conformational averaging is performed[17,33,34], the accuracy of *ab initio* and DFT magnetic parameters cannot be guaranteed. Because vibrational averaging is expensive (at least a third-order response property), this makes the realistically achievable accuracy of the "forward" calculation (geometry → magnetic parameters) unacceptably low. Quite remarkably, this strong geometry response becomes beneficial when the "backward" problem – of recovering the geometry from the experimentally measured magnetic parameters – is considered. This subtle point is illustrated in Figure 2. If the experimental value of a magnetic parameter $m_{\text{exp}}$ is an average with the probability density $p(\vec{r})$ over a thermally accessible convex volume $V$ in the coordinate space

$$m_{\exp} = \int_V m(\vec{r}) p(\vec{r}) dV, \tag{8}$$

then the mean value theorem applies, that is, there exists such a point $\vec{r}_0$ inside that volume that $m(\vec{r}_0) = m_{\exp}$. In other words, *there exists a thermally accessible set of coordinates for which the calculated magnetic parameters would be exactly equal to the full vibrational average*. This point in the coordinate space is hard to identify *a priori*, but it is the point to which a direct structure fitting algorithm would converge.

Similarly, while a minor uncertainty $\Delta r$ in the input coordinates would translate into a large uncertainty $\Delta m$ in magnetic parameters in a "forward" calculation:

$$\Delta m \approx \frac{dm}{dr} \Delta r, \tag{9}$$

the "backward" procedure would gain in accuracy because large deviations in the experimental magnetic parameters would translate into smaller deviations in the extracted coordinates:

$$\Delta r \approx \left(\frac{dm}{dr}\right)^{-1} \Delta m. \tag{10}$$

The conclusion is therefore that the limited accuracy of the DFT calculations of magnetic parameters and the implicit presence of vibrational and conformational averaging in the experimental data do not prevent (and can actually facilitate) structure determination by direct variation of atomic coordinates against the experimental spectrum.

**4. Error functional minimization methods**

Because magnetic resonance spectra have localized sharp features, the least squares error functional defined in Equation (2) is often very hilly – every arrangement of signals in which any two peaks coincide in the theoretical and the experimental spectrum is a local minimum. This situation is illustrated in Figure 3A. Even for small spin systems, unless a very good initial guess is provided, there is little hope of ever finding the global minimum. The use of non-gradient methods (such as simplex, grid search or genetic algorithms) may facilitate fitting, but it does not solve the fundamental problem.

Remarkably, this issue largely disappears if a monotonic function (for example a cumulative integral or a second cumulative integral) is used to represent the spectrum:

$$S(\omega) \rightarrow \int_{-\infty}^{\omega} S(\omega') d\omega' \rightarrow \int_{-\infty}^{\omega} \int_{-\infty}^{\omega'} S(\omega'') d\omega'' d\omega' \tag{11}$$

This is illustrated in Figure 3B – unlike the usual peak representation, the error functional for the least squares fitting of cumulative integrals of the two spectra has a single global minimum. This is not in general the case (counter-examples may be found easily), but using cumulative integral representation does in practice greatly facilitate convergence. This approach was used for the initial steps of the least squares fitting. After convergence is achieved, the optimization is switched to the absorption mode or derivative representation for the final refinement. The use of internal coordinates is also known to facilitate geometry convergence[35] – Z-matrix representation was used for the fitting examples presented below.

In practical fitting runs, the trust-region modification[36] of the L-BFGS quasi-Newton minimization algorithm[30] was found to be superior to conjugate gradient methods and steepest descent, likely a consequence of the quadratic behaviour of the potential energy surface around the energy minimum as well as approximately linear dependence of the magnetic parameters on small coordinate perturbations, which becomes quadratic after the square is taken in Equation (2). Convergence can in practice be further accelerated by supplying a suitably scaled Hessian of the energy (which can be computed at a reasonable cost[37]) as the initial Hessian guess for the minimization of the full error functional in Equation (1).

The fitting restart capability was implemented using intermediate result caching – an MD5 hash of each Gaussian03 DFT job was stored in a database, which was queried before submitting new jobs to avoid the re-calculation of previously computed data points. This is a necessary feature due to the large CPU time cost of computing Equation (4) for every internal coordinate: the calculations reported in this paper consumed over 50,000 hours of CPU time.

## 5. Direct structure fitting to isotropic ESR spectra

The fitting convergence profiles of cyanomethyl (with $^{14}$N isotope for nitrogen), propargyl and tyrosyl radical optimizations performed against their liquid state ESR spectra (digitized from the SDBS-ESR database) are given in Figure 4. DFT calculations were performed using GIAO B3LYP/EPR-II method[38-40] in *Gaussian03*, spin dynamics simulations were done using version 1.0.950 of *Spinach* library[22].

Even in the simple case of propargyl radical, the magnetic parameters calculated at the Born-Oppenheimer energy minimum produced by DFT do not yield a good match to the experimental spectrum (Figure 5) but the minimum of the DSF error functional in Equation (1) does, at the extra energy cost of just 2.78 kJ/mol per internal degree of freedom. The DSF minimum is located very close (0.053 Å RMSD) to the DFT energy minimum. The fitting result does not yield new structural information for the propargyl radical – its structure is known and rigid – but it serves to illustrate the point made in Section 3 about the energy cost of magnetic parameter perturbations: the geometry with the "experimental" magnetic parameters is thermally accessible at room temperature from the Born-Oppenheimer energy minimum geometry. This also illu-strates the mean value theorem argument in Equation (8) – DFT calculations of magnetic parameters are not perfectly accurate due to the limitations of the theory as well as lack of vibrational averaging, but it appears that they are not required to be perfectly accurate.

For the cyanomethyl radical, the DFT energy minimum matches (to within $10^{-3}$ kJ/mol per degree of freedom) the direct structure fitting minimum, but the fitting process shown in Figure 4 (upper panel) was intentionally started from a distorted initial geometry to evaluate the convergence behavior of the L-BFGS minimization algorithm[30] in a situation where the gradient, given in our case by Equation (4), might only have three digits of accuracy. As Figure 4 illustrates, the convergence is quick and robust, with the minimization procedure reaching the minimum of the error functional well before the numerical accuracy of the gradient is exhausted.

The tyrosyl radical (Figure 4, lower panel) has a complicated ESR spectrum that requires the application of the cumulative integral technique to facilitate the fitting. The first round of fitting used the cumulative integral representation to roughly match the extent and the offset of the spectrum. During this stage the energy regularization term in Equation (1) makes sure that the coordinates do not move away from a chemically sensible structure. After a match was achieved on the cumulative integral, the fitting error functional was switched to use the more structured derivative spectrum. Tyrosyl radical also illustrates the mean value argument given in Equation (8) – in liquid state, the experimentally observed isotropic hyperfine couplings of the $CH_2$ protons correspond to the Boltzmann-weighted averages over all possible values of the dihedral angle between the aromatic ring and the backbone fragment, as well as over all vibrational excursions along other coordinates. Still, a point exists in the thermally accessible region of the coordinate space with the "experimental" $CH_2$ hyperfine couplings, and this is the point to which the direct structure fitting algorithm is converging. Other such points may exist and the fitting may therefore need to be repeated from different initial guesses to get the "structure bundle" of the same kind as the ones obtained in protein NMR spectroscopy using molecular dynamics force fields[3-5,7,8].

## 6. Direct structure fitting to powder average ESR spectra

A computationally challenging example of a tyrosyl radical embedded into a protein matrix has been chosen as illustration for the solid state ESR fitting (Figures 6 and 7). Tyrosyl radicals in several ribonucleotide reductase (RNR) enzymes are well characterized, both spectroscopically[19,20] and with X-ray diffraction[41,42]. *Escherichia coli* (PDB: 1RIB/Tyr122) and *Salmonella typhimurium* (PDB: 1R2F/Tyr105) ribonucleotide reductase ESR spectra were chosen for practical demonstration. Atomic coordinate fitting to solid state ESR data requires no modifications to the formalism presented above – powder averaging is a linear procedure and the gradient of the error functional is therefore a powder average of the gradients computed for individual orientations.

The *Escherichia coli* RNR fitting was started from the DFT energy minimum (Figure 6, upper trace) of the neutral radical (*pKa* of the phenoxyl radical is estimated at −2.0[43]) and proceeded in two stages. The initial fit was performed under weak regularization using the absorption mode ESR spectra to locate the offset position and converge on the overall edge-to-edge spectral extent. Because the large hyperfine coupling to the $CH_2$ protons dominates the X-band spectrum, this causes the aromatic ring to rotate into the correct position with respect to the rest of the tyrosine side chain. After the initial convergence was achieved in the absorption mode representation, the fitting was refined in the derivative representation. Three different values of the energy regularization parameter were used, corresponding to the full spectral mismatch being equivalent to $10^4$ kJ/mol (Figure 6, second trace from the top), $10^3$ kJ/mol (third trace) and 100 kJ/mol (fourth trace). The resulting coordinates were compared to the X-ray structures. A nearly perfect spectral fit may be achieved at the total cost of 44.7 kJ/mol in displacement away from the DFT energy minimum, and a very satisfactory fit costs just 6.2 kJ/mol (equivalent to 0.10 kJ/mol per degree of freedom) – well within the room temperature energy budget. The RMSD (carbons only) from the crystal structure for the 0.35 kJ/(mol·df) trace is 0.107 Å. Even the slight bend in the aromatic ring present in the crystal structure is reproduced – the coordinate files are provided in the Supplementary Information.

Similar results were obtained for the *Salmonella typhimurium* RNR (Figure 7) – the DFT energy minimum spectrum is the same as in Figure 6, and a good quality spectral fit may be obtained for a geometry displacement corresponding to 0.13 kJ/mol per degree of freedom. The RMSD (carbons only) from the crystal structure for the 0.13 kJ/(mol·df) trace is 0.106 Å, with the structurally important dihedral angle between the aromatic ring and the side chain reproduced to within 5º – an excellent agreement.

Both fitting runs required considerable, but not astronomical, amounts of CPU time: around 50 L-BFGS iterations with a finite-difference *g*-tensor and HFC gradients computed at each step for 63 coordinates come to a total of 25,200 independent GIAO B3LYP/EPR-II calculations, each taking about half an hour of CPU time on our SGI Altix 4700 supercomputer. The CPU time cost of the same number of spin dynamics simulations (time-domain ESR pulse-acquire experiment with powder averaging on a Lebedev grid) is negligible.

## 7. Open questions and conclusions

Several questions regarding direct structure fitting require further research. We did not evaluate the comparative performance of different exchange-correlation functionals and basis sets, and did not consider any alternatives to DFT. It is very likely that better methods than GIAO B3LYP/EPR-II, both in accuracy and in accuracy-complexity ratio, are available. That having been said, NMR structure determination relies, quite successfully, on molecular dynamics methods – absolute accuracy is clearly not required. Secondly, analytical gradients of magnetic parameters with respect to atomic coordinates are necessary for production-grade structure fitting – the present paper uses expensive finite-difference gradients, and we estimate that the calculations reported could be accelerated by at least two orders of magnitude if analytical gradients were available. Thirdly, the mean value theorem in Equation (8) requires the thermally accessible coordinate volume $V$ to be convex. That is not always the case and the consequences of $V$ being non-convex require further investigation. NMR structure determination uses Monte-Carlo sampling that produces "structure bundles", the same method might work for ESR.

In summary, the computational cost of direct structure fitting to ESR spectra is large, but is no longer prohibitive – at the current development pace, the necessary computing power will reach common desktops in about five years. Large-scale spin dynamics simulations are also becoming possible, meaning that the theoretical modeling cycle may be closed and atomic coordinates iterated directly against the experimental data. The proof-of-principle calculations reported in this work demonstrate the procedure.

**Acknowledgements**

We are grateful to Dima Svistunenko for compiling and sharing in his recent papers[19,20] a database of tyrosyl radical ESR spectra. Helpful discussions with Peter Hore and Hannah Hogben are acknowledged. The project is funded by the EPSRC (EP/F065205/1, EP/H003789/1) and supported by the Oxford e-Research Centre.

**Figure captions**

**Figure 1.** A schematic of the direct structure fitting procedure: molecular coordinates are supplied to a quantum chemistry package, the resulting magnetic parameters are fed to a spin dynamics simulation package (*Spinach*[22] was used in this work) and the squared norm of the deviation from the experimental spectrum is computed. This deviation is iteratively minimized with respect to the molecular coordinates, with a constraint on the molecular energy to stay within the thermally accessible energy range.

**Figure 2.** An example of the improvement in the structure of the fitting error surface produced by the switch from the absorption/emission representation (**A**) to the cumulative integral representation (**B**) of the spectrum. The multiple narrow troughs in the error surface disappear – the cumulative integral representation yields (in this case) a single global minimum that is reachable by any standard minimization procedure from any reasonable starting point in the fitting parameter space. The solid traces in the middle and lower panels show the cross-section of the error surface along the horizontal dotted line.

**Figure 3.** An illustration to the mean value theorem argument used in Equation (8) – the thermal average of a magnetic parameter over a convex volume in the coordinate space has to be equal to the value of that parameter in some (*a priori* unknown) point inside that volume. This is the point to which a direct structure fitting algorithm would converge.

**Figure 4.** Minimization profiles (L-BFGS with cubic line search in internal coordinates) of the error functional in Equation (2) for liquid-state ESR spectra of cyanomethyl (upper panel), propargyl (middle panel) and tyrosyl (lower panel) radicals. The axis units are arbitrary.

**Figure 5.** Direct structure fitting result for the liquid state spectrum of the propargyl radical. Complete agreement with the experimental data was obtained at the energy cost of 2.78 kJ/mol per internal degree of freedom brought about by a geometry change (0.053 Å RMSD from the DFT energy minimum).

**Figure 6.** Direct structure fitting results for the solid-state ESR spectrum of tyrosyl radical in *Escherichia coli* ribonucleotide reductase (PDB: 1RIB/Tyr122). The DFT energy minimum (top trace) is very far from both the X-ray geometry and the experimental ESR spectrum (bottom trace). The three direct structure fitting traces differ in the value of the energy regularization parameter – the final energy cost of spectral fit

shown to the left of each trace. The coordinates corresponding to the middle trace are within 0.107 Å RMSD (carbons only) from the crystal structure geometry.

**Figure 7.** Direct structure fitting results for the solid-state ESR spectrum of tyrosyl radical in *Salmonella typhimurium* ribonucleotide reductase (PDB: 1R2F/Tyr105). The DFT energy minimum (Figure 6, top trace) is very far from both the X-ray geometry and the experimental ESR spectrum (bottom trace). The three direct structure fitting traces differ in the value of the energy regularization parameter – the final energy cost of spectral fit shown to the left of each trace. The coordinates corresponding to the top trace are within 0.106 Å RMSD (carbons only) from the crystal structure geometry.

**Table 1.** Derivatives of isotropic hyperfine couplings with respect to the internal coordinates of the allyl radical, in units of Gauss/Ångstrom for bonds and Gauss/radian for bond angles (B3LYP/EPR-II calculation using four-point central finite differences at the energy minimum in the gas phase).

| Internal coordinate | $^{13}C^{(1)}$ | $^{1}H^{(1)}$ | $^{1}H^{(1')}$ | $^{13}C^{(2)}$ | $^{1}H^{(2)}$ | $^{13}C^{(3)}$ | $^{1}H^{(3)}$ | $^{1}H^{(3')}$ |
|---|---|---|---|---|---|---|---|---|
| $R[C^{(1)}-H^{(1)}]$ | 22.1 | –11.6 | –0.9 | 0.0 | 0.5 | 0.6 | –0.6 | –0.5 |
| $R[C^{(1)}-H^{(1')}]$ | 22.8 | –1.1 | –13.0 | –0.1 | 0.7 | 1.1 | 0.1 | –0.2 |
| $R[C^{(1)}-C^{(2)}]$ | 57.1 | –27.4 | –26.2 | –15.7 | 4.0 | –26.5 | 22.2 | 22.1 |
| $R[C^{(2)}-H^{(2)}]$ | 0.2 | –0.1 | –0.3 | –10.6 | 2.9 | 0.2 | –0.1 | –0.3 |
| $R[C^{(2)}-C^{(3)}]$ | –26.5 | 22.2 | 22.1 | –15.7 | 4.0 | 57.1 | –27.4 | –26.2 |
| $R[C^{(3)}-H^{(3')}]$ | 1.1 | 0.1 | –0.2 | –0.1 | 0.7 | 22.8 | –1.1 | –13.0 |
| $R[C^{(3)}-H^{(3)}]$ | 0.6 | –0.6 | –0.5 | 0.0 | 0.5 | 22.1 | –11.6 | –0.9 |
| $A[H^{(1)}-C^{(1)}-H^{(1')}]$ | –0.63 | 0.20 | –2.56 | –0.06 | –0.33 | –0.05 | 0.02 | –0.46 |
| $A[H^{(1')}-C^{(1)}-C^{(2)}]$ | 0.16 | 2.55 | –3.17 | –2.00 | 0.11 | –0.03 | 0.12 | –0.32 |
| $A[C^{(1)}-C^{(2)}-H^{(2)}]$ | 1.03 | –1.82 | –0.70 | 0.00 | 0.00 | –1.03 | 1.82 | 0.70 |
| $A[C^{(1)}-C^{(2)}-C^{(3)}]$ | 0.25 | 0.10 | –0.94 | –4.77 | 0.90 | –0.83 | 2.12 | –0.42 |
| $A[C^{(2)}-C^{(3)}-H^{(3)}]$ | 0.05 | –0.02 | 0.46 | 0.06 | 0.33 | 0.63 | –0.20 | 2.56 |
| $A[C^{(2)}-C^{(3)}-H^{(3')}]$ | 0.03 | 0.31 | 0.04 | –1.94 | 0.42 | 0.79 | 2.21 | –0.58 |

**Table 2.** Affordability of isotropic hyperfine coupling perturbation, in units of Gauss/(kJ/mol), along specific internal coordinates of the allyl radical.

| Internal coordinate | $^{13}C^{(1)}$ | $^{1}H^{(1)}$ | $^{1}H^{(1')}$ | $^{13}C^{(2)}$ | $^{1}H^{(2)}$ | $^{13}C^{(3)}$ | $^{1}H^{(3)}$ | $^{1}H^{(3')}$ |
|---|---|---|---|---|---|---|---|---|
| R[C$^{(1)}$–H$^{(1)}$] | 1.33 | –0.70 | –0.05 | 0.00 | 0.03 | 0.04 | –0.04 | –0.03 |
| R[C$^{(1)}$–H$^{(1')}$] | 1.37 | –0.06 | –0.78 | 0.00 | 0.04 | 0.06 | 0.01 | –0.01 |
| R[C$^{(1)}$–C$^{(2)}$] | 2.99 | –1.43 | –1.37 | –0.82 | 0.21 | –1.39 | 1.16 | 1.16 |
| R[C$^{(2)}$–H$^{(2)}$] | 0.01 | –0.01 | –0.02 | –0.65 | 0.18 | 0.01 | –0.01 | –0.02 |
| R[C$^{(2)}$–C$^{(3)}$] | –1.39 | 1.16 | 1.16 | –0.82 | 0.21 | 2.99 | –1.43 | –1.37 |
| R[C$^{(3)}$–H$^{(3')}$] | 0.04 | –0.04 | –0.03 | 0.00 | 0.03 | 1.33 | –0.70 | –0.05 |
| R[C$^{(3)}$–H$^{(3)}$] | 0.06 | 0.01 | –0.01 | 0.00 | 0.04 | 1.37 | –0.06 | –0.78 |
| A[H$^{(1)}$– C$^{(1)}$– H$^{(1')}$] | –0.12 | 0.04 | –0.50 | –0.01 | –0.07 | –0.01 | 0.00 | –0.09 |
| A[H$^{(1')}$– C$^{(1)}$– C$^{(2)}$] | 0.03 | 0.45 | –0.56 | –0.36 | 0.02 | –0.01 | 0.02 | –0.06 |
| A[C$^{(1)}$– C$^{(2)}$– H$^{(2)}$] | 0.18 | –0.31 | –0.12 | 0.00 | 0.00 | –0.18 | 0.31 | 0.12 |
| A[C$^{(1)}$– C$^{(2)}$– C$^{(3)}$] | 0.03 | 0.01 | –0.13 | –0.65 | 0.12 | –0.11 | 0.29 | –0.06 |
| A[C$^{(2)}$– C$^{(3)}$– H$^{(3)}$] | 0.01 | 0.00 | 0.09 | 0.01 | 0.07 | 0.12 | –0.04 | 0.51 |
| A[C$^{(2)}$– C$^{(3)}$– H$^{(3')}$] | 0.01 | 0.06 | 0.01 | –0.38 | 0.08 | 0.16 | 0.43 | –0.11 |

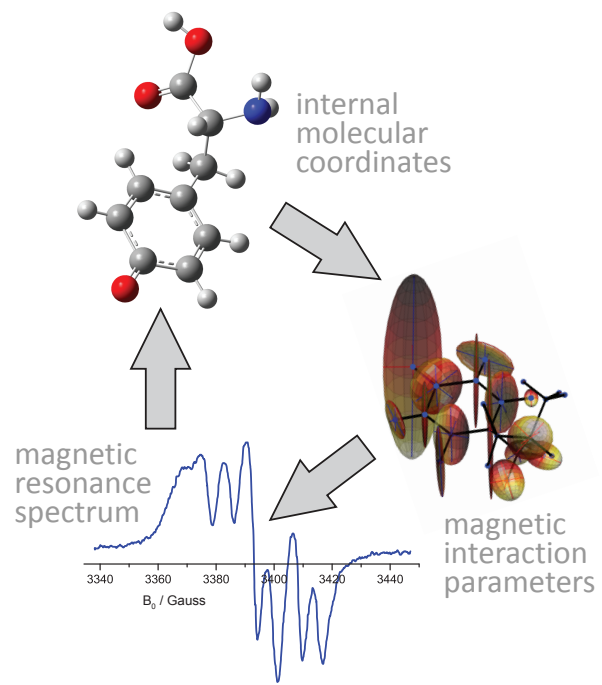

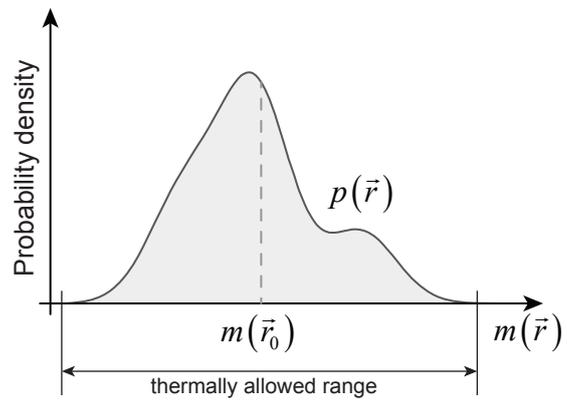

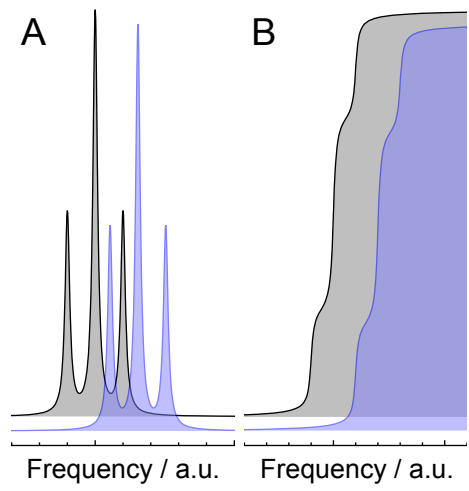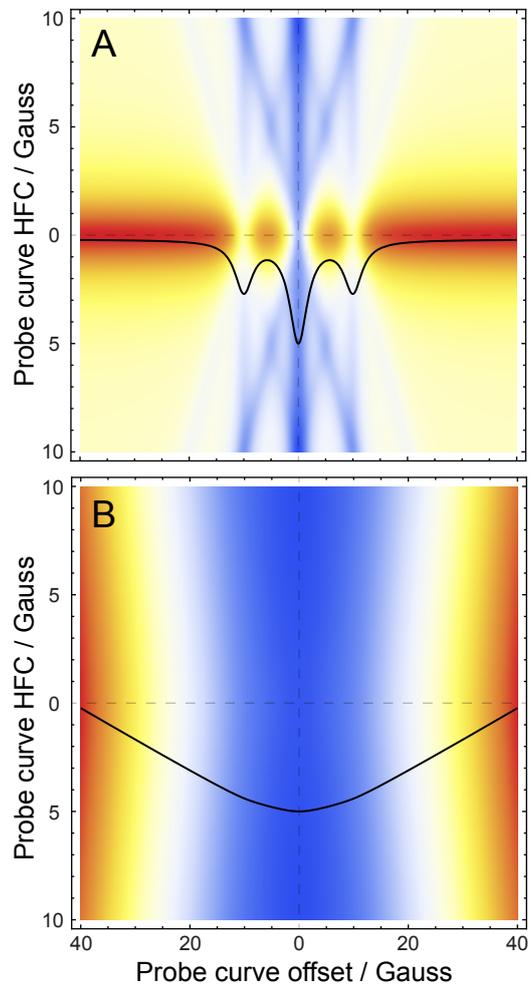

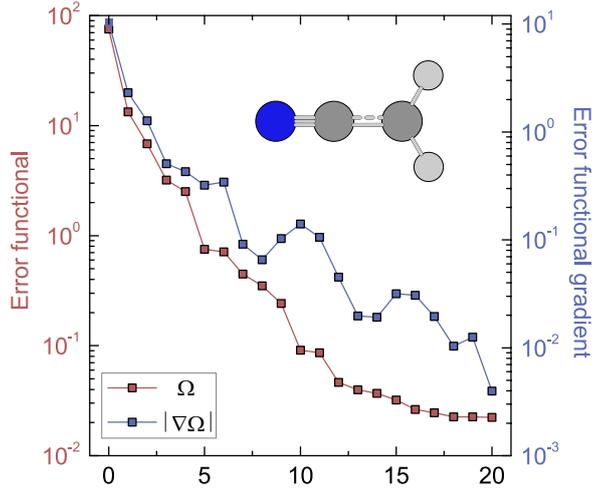
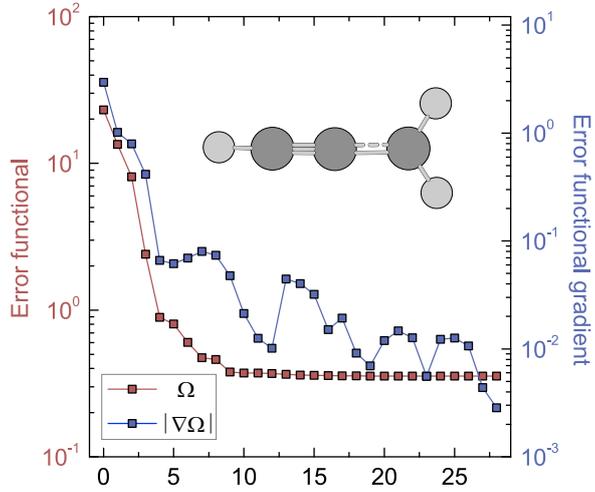
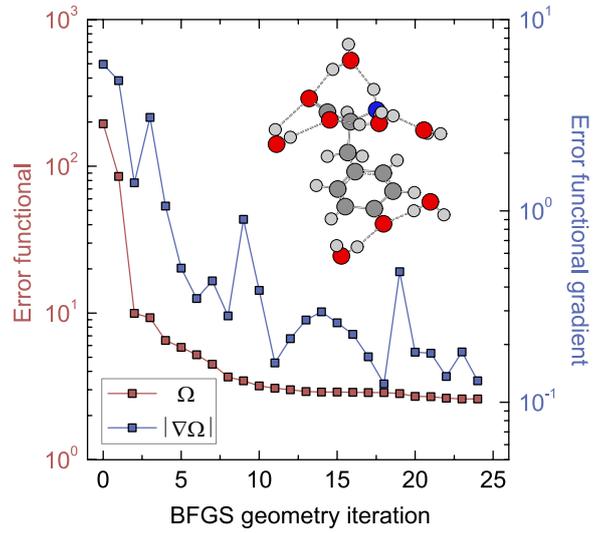

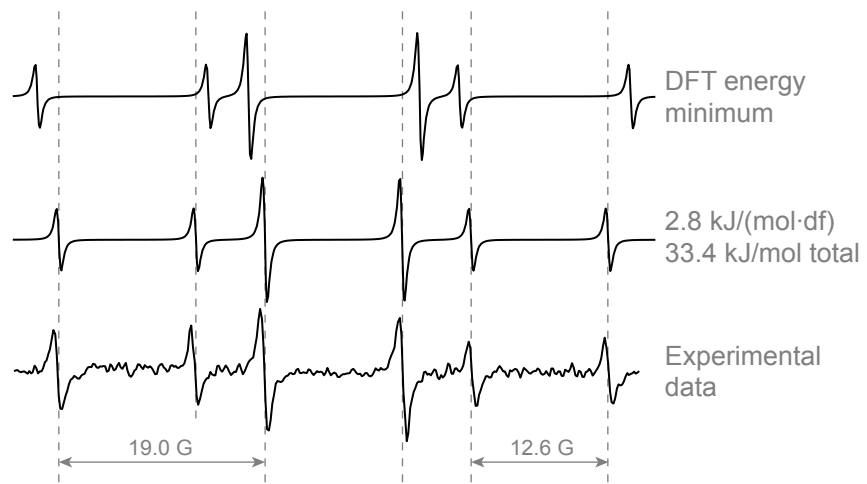

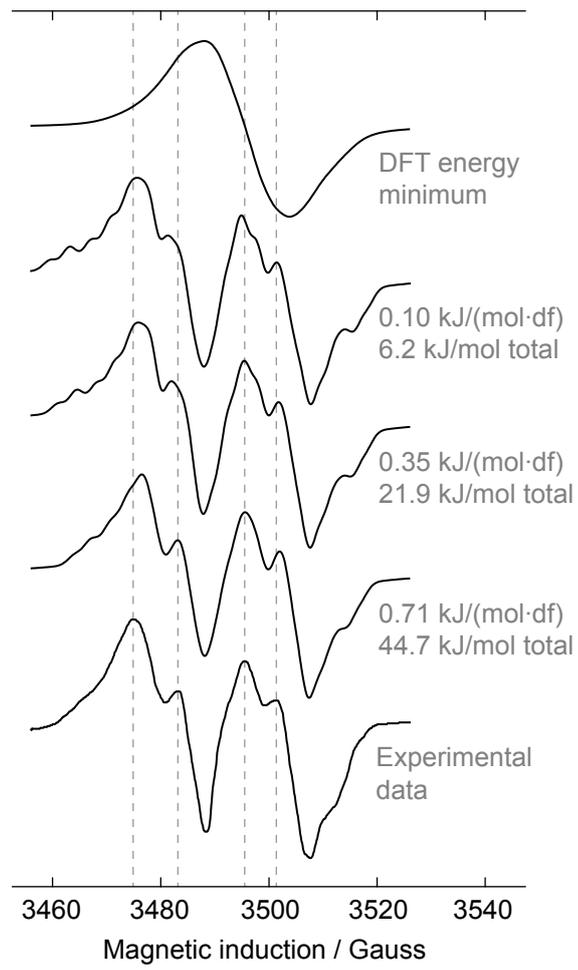

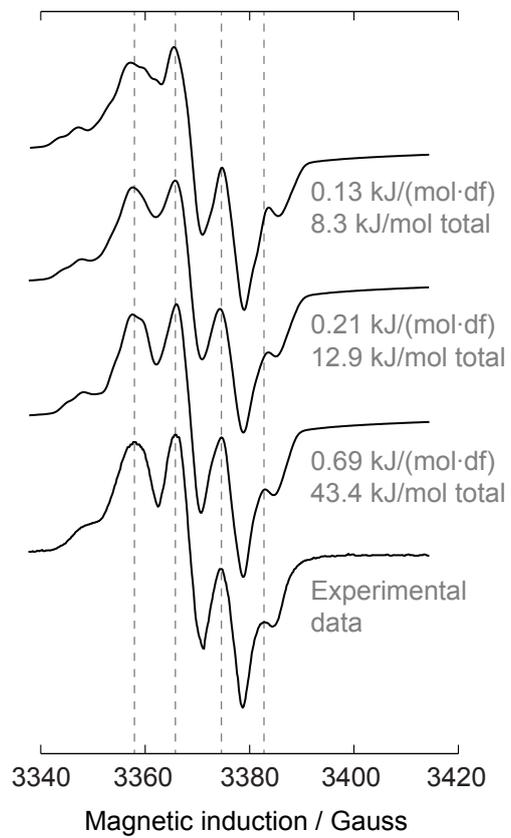